\documentclass[sn-basic]{sn-jnl}
\usepackage{graphicx}
\usepackage{amsmath,amsfonts}
\usepackage{amsbsy}
\usepackage{bm}
\usepackage{graphics}

\begin{document}

\title[Skin Anisotropy Analysis with Elastic Waves and Baysian Modelling]{Analysis of in-vivo skin anisotropy using elastic wave measurements and Bayesian modelling}

\author[1,2]{Matt Nagle\textsuperscript{*}}
\author[2]{Susan Price}
\author[2]{Antonia Trotta}
\author[5,2]{Michel Destrade}
\author[3]{Michael Fop\textsuperscript{*$\dagger$}}
\author[2,4]{Aisling Ní Annaidh\textsuperscript{$\dagger$}}

\affil[1]{SFI Centre for Research Training in Foundations of Data Science, University College Dublin, Belfield, Dublin 4, Ireland}
\affil[2]{School of Mechanical and Materials Engineering, University College Dublin, Belfield, Dublin 4, Ireland}
\affil[3]{School of Mathematics and Statistics, University College Dublin, Belfield, Dublin 4, Ireland}
\affil[4]{Charles Institute of Dermatology, University College Dublin, Belfield, Dublin 4, Ireland}
\affil[5]{School of Mathematical and Statistical Sciences, University of Galway, Galway, Ireland}

\footnotetext[1]{Corresponding authors: matt.nagle@ucdconnect.ie; michael.fop@ucd.ie;}
\footnotetext[2]{Contributing authors: susan.price@ucdconnect.ie; antonia.trotta@ucdconnect.ie; michel.destrade@universityofgalway.ie; aisling.niannaidh@ucd.ie;}
\footnotetext[3]{These authors contributed equally to this work.}

\newpage
 
\abstract{In vivo skin exhibits viscoelastic, hyper-elastic and non-linear characteristics. It is under a constant state of non-equibiaxial tension in its natural configuration and is reinforced with oriented collagen fibers, which gives rise to anisotropic behaviour. Understanding the complex mechanical behaviour of skin has relevance across many sectors including pharmaceuticals, cosmetics and surgery. However, there is a dearth of quality data characterizing the anisotropy of human skin in vivo. The data available in the literature is usually confined to limited population groups and/or limited angular resolution. Here, we used the speed of elastic waves travelling through the skin to obtain measurements from 78 volunteers ranging in age from 3 to 93 years old. Using a Bayesian framework allowed us to analyse the effect that age, gender and level of skin tension have on the skin anisotropy and stiffness. First, we propose a new measurement of anisotropy based on the eccentricity of angular data and conclude that it is a more robust measurement when compared to the classic ``anisotropic ratio". Our analysis then concluded that in vivo skin anisotropy increases logarithmically with age, while the skin stiffness increases linearly along the direction of Langer Lines. We also concluded that the gender does not significantly affect the level of skin anisotropy, but it does affect the overall stiffness, with males having stiffer skin on average. Finally, we found that the level of skin tension significantly affects both the anisotropy and stiffness measurements employed here. This indicates that elastic wave measurements may have promising applications in the determination of in vivo skin tension. In contrast to earlier studies, these results represent a comprehensive assessment of the variation of skin anisotropy with age and gender using a sizeable dataset and robust modern statistical analysis. This data has implications for the planning of surgical procedures and questions the adoption of universal cosmetic surgery practices for very young or elderly patients.}


\maketitle

\section{Introduction}\label{sec1}
The skin is a vital organ for a range of bodily functions including protection from the environment and temperature regulation \citep{Joodaki2018}. It is constantly under varying amounts of tension and must be able to withstand significant flexion and deformation for daily tasks like locomotion. Understanding the mechanical properties of the skin is important for many different applications and industries: in the cosmetic industry, products must be assessed in terms of emolliency and hydration of the skin; in the design of anthropomorphic devices like crash test dummies and surgical simulators \citep{Joodaki2018}, the mechanical behaviour of the skin must be replicated as closely as possible; and in a surgical setting, a thorough understanding of the skin's mechanical properties is essential. For example, understanding skin growth through tissue expansion is necessary for breast reconstruction and burn victims \citep{Pamplona2014}.

Previous publications have examined many different mechanical properties of skin including viscoelasticity \citep{Ruvolo2007}, the nonlinear stress-strain relationship \citep{Maurel1998, PaillerMattei2008}, failure properties \citep{Dombi1993, NiAnnaidh2012, Ottenio2015} and anisotropy \citep{NiAnnaidh2012, Khatyr2004}. The experimental methods employed have included extension \citep{Khatyr2004}, suction \citep{Hendriks2006}, torsion\textsuperscript{1}, indentation \citep{PaillerMattei2008} and expansion \citep{Pamplona2014}, amongst others. In this paper, we focus mainly on the anisotropic nature of the in vivo skin tension which was first noted in the 19th century by anatomist Karl Langer \citep{Langer1978}. Preferred lines of tension have become known as "Langer Lines" or skin tension lines and are used by surgeons to select the optimum orientation of skin incisions so as to reduce scarring \citep{Paul2018}. Identification of these patient-specific lines is non-trivial and surgeons must rely on generic maps or an imprecise pinch test that requires significant experience to interpret \citep{Deroy2017, Seo2013, Destrade2019}. Recent research has shown that minimising the skin tension across wounds is the single most important factor in scar prevention that is within a surgeon's control \citep{Son2014, Stowers2021}. To that end, quantitative knowledge of both the direction and the anisotropic nature of skin tension lines is an essential component of wound closure. A deeper understanding of how they vary across a population may provide further optimisation of closure techniques, particularly for elderly or very young patient cohorts.

More recently, researchers have sought to develop techniques to determine the in vivo orientation of skin tension lines objectively: these include those using suction based devices \citep{Laiacona2019}, in vivo extensometry \citep{Paul2017, Paul2018} and elastic wave propagation \citep{Deroy2017, Ruvolo2007}. These papers have shown that these techniques can successfully identify skin tension lines, and that their orientation is patient-specific, depending on many different factors including location, age, health, BMI, ethnicity and hydration \citep{Deroy2017}. 
However, with the exception of \cite{Ruvolo2007}, none of these papers have comprehensively considered the level of anisotropy of these skin tension lines in vivo, i.e. how the tension levels in two orthogonal directions differ, and how that aspect varies by age and gender. In \citep{Ruvolo2007}, 239 volunteers ranging in age from newborn to 75 years old underwent testing using elastic wave propagation. \cite{Ruvolo2007} noted that previous studies had found only a weak dependence on elastic wave speed with age \citep{Dahlgren1984, Vexler1999, HermannsLe2001}; however, these studies all employed poor resolution angular data (measurements taken every 45 degrees) and, having undersampled, they may have missed important information. While great care was taken by \cite{Ruvolo2007} to overcome this issue by sampling every 3 degrees, their anisotropy results are reported in terms of the classic ``anisotropic ratio", which is a simple ratio between the fastest and slowest wave speed. In the current study we propose, instead, to report the eccentricity of an ellipse fit to the circular data, which may offer a more representative and robust measure of the in vivo anisotropy. Additionally, the current study includes a sizeable dataset with a large range of ages (78 individuals, age 3-93) in contrast to \cite{Laiacona2019} (19 subjects, age 18-30), \cite{Boyer2009} (20 subjects, age 20-65), and finally, \cite{HermannsLe2001} (110 subjects, age 19-93), who did not include infants.

Finally, previous papers have all employed hypothesis testing to support their conclusions. It is now commonly accepted that there are significant issues with the use of p-values in scientific research \citep{Wasserstein2016}. Bayesian methods for data analysis provide a principled framework for inference, uncertainty assessment and inclusion of prior information \citep{Gelman1995}. These methods are flexible, capable of handling complex correlation structures and can eliminate the need for p-values and Null Hypothesis Significance Testing (NHST) \citep{Kruschke2010, Dunson2001, Schoot2021}.

The objective of this paper is to determine the level of in vivo skin anisotropy and determine how it varies with age and gender. Here we used elastic wave propagation to determine the speed of surface waves traveling through the skin of 72 subjects (age 3-93). Bayesian statistical methods were then employed to examine the significance and effects of age and gender. Furthermore, we examine how skin anisotropy is affected by skin tension and discuss its implications for surgical practice.

\section{Materials and Methods}\label{sec2}

\subsection{Data Collection}\label{subsec2}
\begin{figure}[h]
\centering
\includegraphics[width=0.7\textwidth]{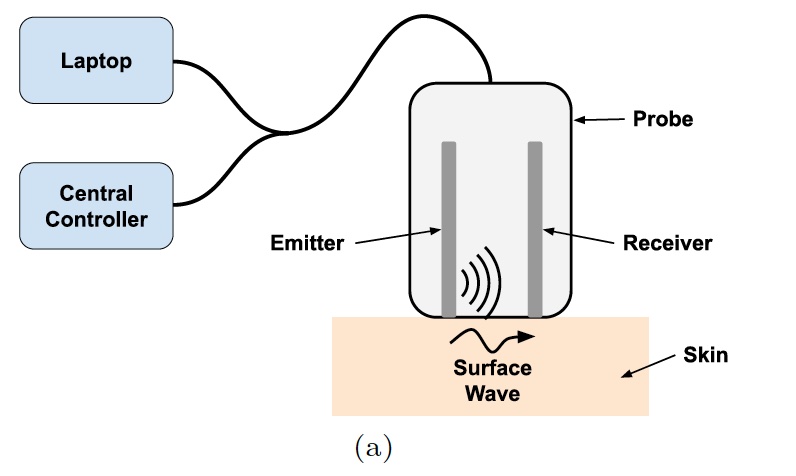} 
\includegraphics[width=0.7\textwidth]{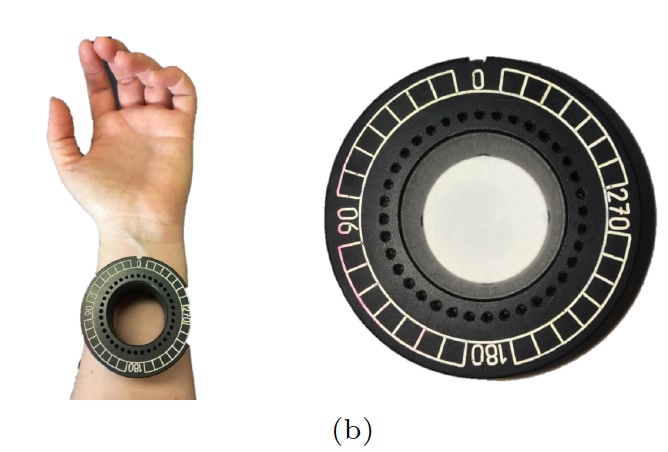} 
\caption{Experimental set up displaying (a) the laptop, central controller and Reviscometer probe (b) Measuring site set up and plastic probe fixture to facilitate accurate angular measurements.}
\label{fig:1}
\end{figure}

The Reviscometer\textregistered{} (Model RVM 600, Courage \& Khazaka Electronic GmbH) is used to examine the mechanical properties of the skin. The device consists of a handheld probe connected to a central controller and a laptop (see Figure \ref{fig:1}a). The tip of the probe contains two piezoelectric transducers that are 2 mm apart. One transducer emits a Rayleigh surface wave in the form of an acoustic pulse on the skin surface, the other detects the resulting wave and records the time taken for the wave to propagate across the surface of the skin, in one orientation. A hollow plastic fixture also facilitates precise measurements at $10^{\circ}$ increments (see Figure \ref{fig:1}b) allowing us to see how the surface wave speed varies for different angles. By default, the measurement is in arbitrary units called ``Resonance Running Time" (RRT). The device was calibrated by assuming the wave speed follows that of a Rayleigh wave travelling on an unstressed, incompressible, linear elastic, isotropic material \citep{Liang2009}. The wave speed is then related to the stiffness through:

\begin{equation}
E=\rho v^{2}(3.284),
\label{eq:1}
\end{equation}
where $E$ is the Young modulus, $\rho$ is the material density and $v$ is the wave speed \citep{Bayon2005}. Specifically, the Young moduli of 3 elastomers (Techsil 25 Silicone, Polyurethane and MVQ Elastomer) were determined with tensile tests and the average RRT (3 tests) determined for each sample. The conversion for each material is detailed in Table \ref{tab:1}. The average RRT was found to be $0.284~\mu$s.

Ethical approval for the study was granted by the Human Research Ethics Commitee at University College Dublin (25-18-75). A total of 78 subjects were tested with 37 female and 41 male volunteers, aged between 3 and 92 years of age, see Table \ref{tab:2}.

Measurements were obtained on either the left or right volar forearm approximately 5 cm proximal to the wrist, see Figure \ref{fig:1}b. This site was chosen as a convenient, flat surface with minimal body hair, veins or bone. For each volunteer, two configurations were explored: the ``natural configuration", where measurements were carried out on the skin with no interference and the ``stretched configuration", where an additional stretch was applied to the skin in the direction of the fastest traveling surface wave. The purpose of this protocol was to demonstrate that an increase in skin tension corresponds to an increase in the wave speed (or equivalently, a decrease in the arrival time) which can be measured in vivo by the Reviscometer\textsuperscript{\textregistered{}}. Measurements were taken from $0^{\circ}-360^{\circ}$ in $10^{\circ}$ increments giving a total of 36 observations. This method was repeated three times per volunteer and an average was calculated\footnote{For practical reasons, only one set of observations was performed on younger subjects who found it difficult to remain in the position required for the study.}.

\begin{table}[h]
\centering
\caption{Analytical conversions for three different materials from RRT to seconds.}
\begin{tabular}{|l||c|c|c|}
\hline
Elastomer Material & $\rho(\text{kg/m}^2)$ & E Avg. RRT & 1 RRT Conversion ($\mu$s) \\
\hline
Techsil 25 Silicone & 928 & 463260 503 & 0.322 \\
Polyurethane & 1237 & 2517780 323 & 0.249 \\
MVQ Elastomer & 1348 & 330000 820 & 0.282 \\
\hline
Average & & & 0.284 \\
\hline
\end{tabular}
\label{tab:1}
\end{table}

\begin{table}[h]
\centering
\caption{Age Distribution of Subjects in 10 year increments.}
\begin{tabular}{|l||l|l|l|l|}
\hline
Age Range & Total Number  & Number  & Mean Age & Standard Deviation \\
(years) & of Subjects &  of Females & (years) & (years) \\
\hline
0-10 & 7 & 5 & 5.9 & 2.73 \\
11-20 & 3 & 10 & 16.1 & 3.07 \\
21-30 & 18 & 8 & 25.6 & 2.28 \\
31-40 & 9 & 2 & 34.3 & 2.17 \\
41-50 & 4 & 7 & 45.3 & 3.04 \\
51-60 & 10 & 5 & 54.3 & 2.41 \\
61-70 & 2 & 7 & 64.4 & 2.30 \\
71-80 & 2 & 3 & 73.7 & 1.53 \\
81-90 & 6 & 4 & 85.2 & 2.79 \\
90+ & 1 & 1 & 92 & \\
\hline
\end{tabular}
\label{tab:2}
\end{table}

Using the ``natural configuration" data, the direction of the Langer line was identified to the nearest $10^{\circ}$ by finding the direction in which the shortest arrival time was recorded. Then, using surgical tape, the skin was stretched in the direction of the identified Langer line. The ``stretched configuration" test was then repeated a further three times and an average was calculated.

\subsection{Anisotropy Measurement}\label{subsec2:2}
As discussed in Section \ref{sec1}, there is a need to quantify skin anisotropy and understand its relationship to skin tension. A number of previous studies have been performed where measures of skin anisotropy are calculated. The most commonly used measure is the ratio of the maximum and the minimum measured value (arrival time or wave speed) \citep{Vexler1999, Ruvolo2007, Ohshima2011, Deroy2017}. In our study, using the Reviscometer, this Anisotropic Ratio (AR) is the ratio of the maximum and minimum RRT values:
\begin{equation}
AR=\frac{RRT_{max}}{RRT_{min}}.
\label{eq:2}
\end{equation}
While this measure can be indicative of the degree of anisotropy, it is also very sensitive to outliers in the data. Furthermore, if measurements are taken from $0^{\circ}-360^{\circ}$ (as is often the case), this measurement ignores much of the available data and uses only the maximum and minimum values.

\begin{figure}[h]
\centering
\includegraphics[width=0.7\textwidth]{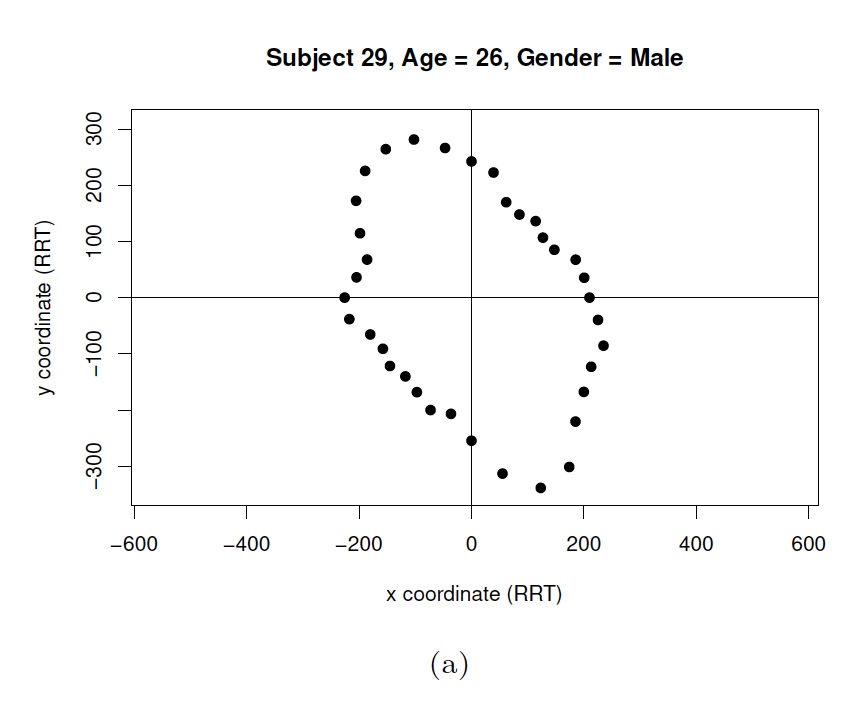} 
\includegraphics[width=0.7\textwidth]{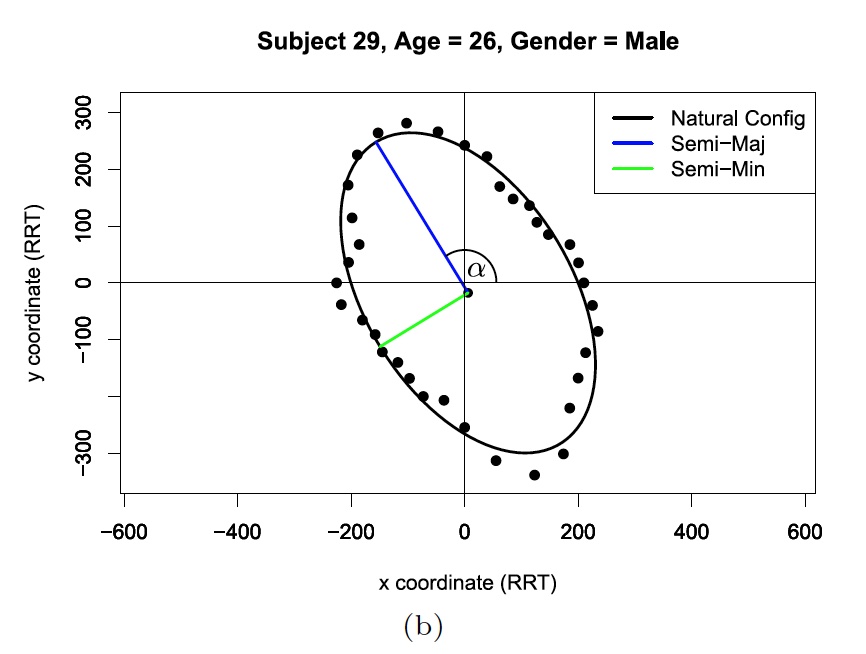} 
\caption{Visualisation of (a) typical raw Reviscometer data from a 26-year-old male subject and (b) the fit ellipse. Note that the Euclidean distance from the origin is the arrival time of the surface wave in units of RRT at that angle (measured counter-clockwise from the positive side of the x-axis).}
\label{fig:2}
\end{figure}

In this paper we suggest an alternative measure of anisotropy that is less susceptible to individual outliers and considers all measurements from $0^{\circ}-360^{\circ}$. We consider all RRT observations and fit an ellipse to them. The eccentricity of this fit ellipse is indicative of the material anisotropy. Assuming an ellipse is an appropriate model for our data, we plot the raw data by allowing the arrival time to be the distance from the origin and the angle to be the angle of inclination from the positive side of the x-axis, see Figure \ref{fig:2}a. Using this representation in Cartesian coordinates, an ellipse can be fit to the raw data using the least squares approach detailed in \cite{Fitzgibbon1999}, which is implemented in the function ``EllipseDirectFit" from the R package ``conicfit" \citep{Gama2015, RCoreTeam2020}, see Figure \ref{fig:2}b. We can then extract the geometric parameters from the ellipse and use them to infer real-world attributes of the skin. All code used can be found in the public GitHub repository accompanying this paper \url[{https://github.com/matt-nagle/Analysis-of-in-vivo-skin-anisotropy-using-elastic-wave-measurements-and-Bayesian-modelling}].

The geometric parameters extracted from each ellipse were: the lengths of the semi-major and semi-minor axes and the angle between the semi-major axis and the positive x-axis (tilt angle). The tilt angle of the ellipse provides the direction of the slowest traveling wave and $90^{\circ}$ from this is the fastest traveling wave which corresponds to the direction of Langer lines \citep{Deroy2017}. The lengths of the semi-major and semi-minor axes allow us to calculate both the area, $A$, and the eccentricity, $e$, of the fit ellipse using Equations \eqref{eq:3} and \eqref{eq:4} respectively:
\begin{equation}
{A} = \pi ab,
\label{eq:3}
\end{equation}
\begin{equation}
e=\sqrt{1-\frac{b^{2}}{a^{2}}}
\label{eq:4}
\end{equation}
where $a$ is the length of the semi-major axis and $b$ is the length of the semi-minor axis. The area relates to an average measure of arrival time in all directions. The smaller the area, the faster all waves are traveling on average. Following Equation \eqref{eq:1}, we can relate this wave speed directly to skin stiffness. This parameter is independent of the anisotropic nature of the measurements.

Eccentricity relates to the material anisotropy; an eccentricity of 0 indicates a circle, i.e. the speed of the wave does not vary depending on the angle and the material is perfectly isotropic. As the eccentricity increases from 0, the difference between the slowest wave and the fastest wave also increases, i.e. the skin demonstrates more and more anisotropy. An eccentricity of 1 indicates a straight line which is perfectly transversely isotropic. In practice, eccentricity values in our study mostly fell between 0.5 and 0.9.

\subsection{Simulation Study}\label{subsec2:3}
To evaluate the performance of the two different measures of anisotropy ($AR$ vs $e$) a simulation study was performed. Simulation studies are computer-based experiments that use artificially generated data to examine the performance of different methods. Knowledge of the underlying data generation mechanism enables a thorough evaluation and comparison \citep{Morris2019}.

\begin{figure}[h]
\centering
\includegraphics[width=0.9\textwidth]{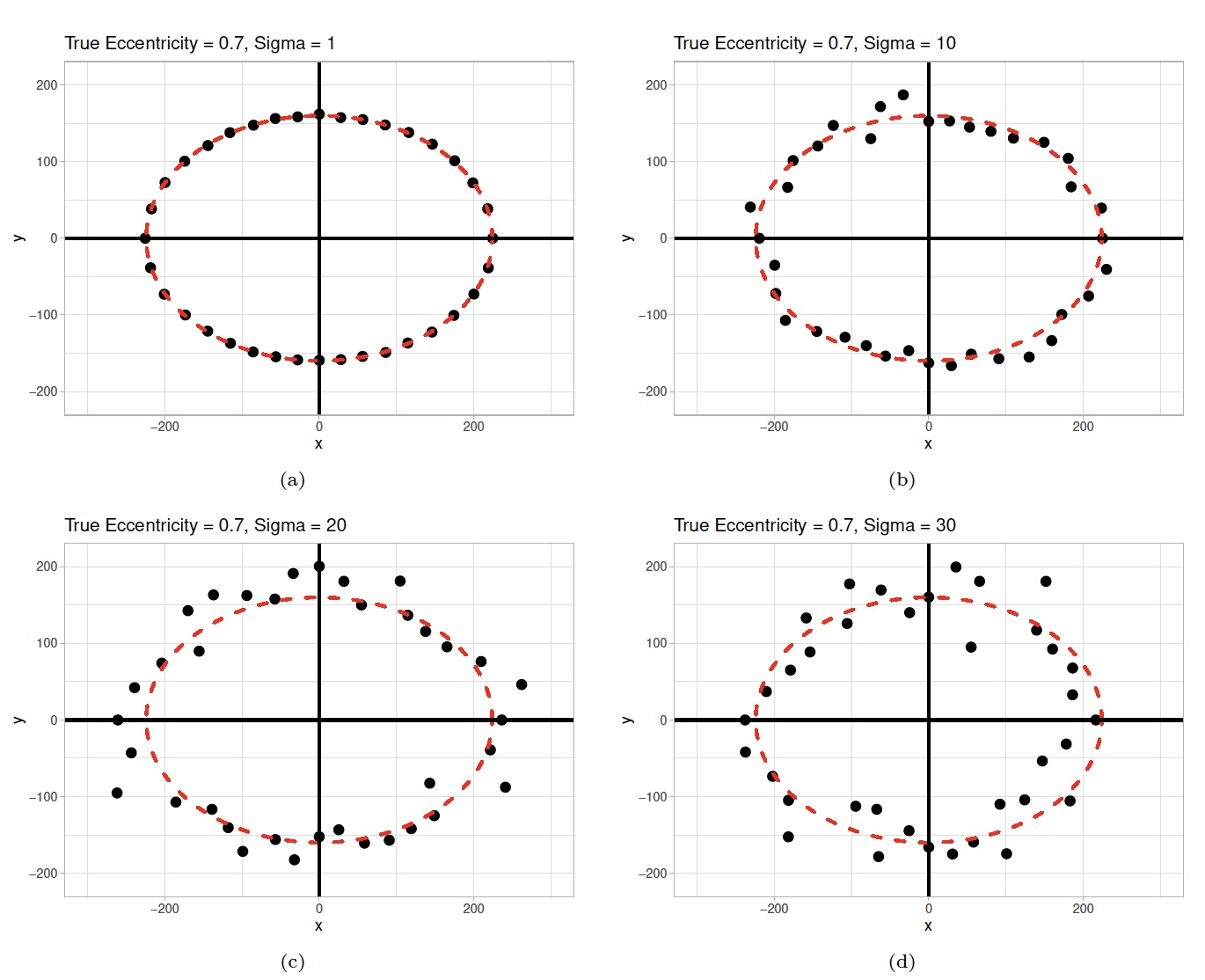} 
\caption{Sample simulated data for an eccentricity of 0.7 showing (a) essentially no noise, $\sigma=1$, (b) low noise, $\sigma=10$, (c) medium noise, $\sigma=20$, (d) high noise, $\sigma=30$. The dashed red line is the shape of the underlying ellipse before the noise was added.}
\label{fig:3}
\end{figure}

In short, simulated data was generated following a known regular shape, random noise was added to the data, then the two measures of anisotropy were compared to the known true values. Ellipses with a fixed value of 160 RRT for the semi-minor axis with eccentricities $e=$ [0.5, 0.7, 0.9] were selected as being representative of our dataset. Noise was added to the ellipses using a random draw from a normal distribution with mean 0 and standard deviation $\sigma$. Four different values of $\sigma$ were used, ranging from very low to high amounts of noise, $\sigma=[1,10,20,30]$, see Figure \ref{fig:3}.

For each set of points, both measures of anisotropy (eccentricity of the fit ellipse and the AR) were calculated and stored. This procedure was performed 10,000 times for each value of $\sigma$.

\subsection{Bayesian Data Analysis}\label{subsec2:4}
As discussed in Section \ref{sec1}, Bayesian methods for data analysis are often appealing as they avoid some of the potential issues with p-values and NHST \citep{Wasserstein2016}. In general, in a frequentist approach, a model coefficient is a single deterministic fixed value. On the other hand, in a Bayesian framework, model coefficients are random quantities which are assumed to have probability distributions that convey prior beliefs and the uncertainty around their value. The aim is to perform inference on the ``posterior distribution" of the parameters, taking into account both the prior knowledge and the evidence provided by the observed data. Inference in this setting is typically performed using Markov Chain Monte Carlo (MCMC) methods \citep{Gelman1995}, which are employed to derive a large chain of estimates for each model coefficient. Each coefficient estimate in the chain is a draw from the posterior distribution and considering a large number of estimates gives us the shape of this distribution. This allows us to quantify not only the mean value of the coefficient (``posterior mean") but also the uncertainty we have in each coefficient by considering the spread of the distribution. In contrast to the confidence interval in the frequentist approach, Highest Posterior Density Intervals (HPDI) can be defined which directly relate to a probability, i.e. there is a probability of 0.95 that the true value of the coefficient lies within the 0.95 HPDI. Thus, if we are interested in covariate significance, we can simply examine the posterior distribution of the corresponding coefficient. For example, if 0 is within the 0.95 HPDI for a certain coefficient then we do not have enough evidence to suggest that the coefficient is significantly different from zero and must conclude the covariate does not have a significant effect on our outcome variable. Bayesian approaches are often much more flexible and allow for greater model complexity; this is especially important for the current application as we need to account for correlated target variables, as well as circular target variables which would be very difficult to do with a frequentist approach \citep{Gelman1995}.

To examine the influence that the subject age, subject gender and the applied additional stretch have on the skin properties anisotropy, average stiffness and stiffness in the direction of the Langer line (measured by the eccentricity, area and length of the semi-minor axis of the fit ellipse respectively), a Bayesian multivariate outcome regression model was built. The multivariate approach was selected to account for the fact that $e$ and $A$ are both calculated using the semi-major and semi-minor axes and are therefore correlated. The model is of the form:

\begin{equation}
\begin{pmatrix}
\text{log(Eccentricity)} \\
\text{Area/1000} \\
\text{Semi-minor Axis}
\end{pmatrix}
= \mathbf{a} + \mathbf{B}
\begin{pmatrix}
\text{Age} \\
\text{Gender} \\
\text{Config.}
\end{pmatrix}
+ \mathbf{E},
\label{eq:5}
\end{equation}
where $\mathbf{a}$ is a three-dimensional vector representing the intercept, $\mathbf{B}$ is a $3\times3$ matrix of coefficients for age, gender and configuration, and $\mathbf{E}\sim N(0,\Sigma)$ is the error term. Note that as the eccentricity is a parameter bounded by 0 and 1, a regular linear model would not be suitable as we cannot obtain values of $e>1$ or $e<0$ regardless of the input values. Therefore, a log of the eccentricity was taken as the outcome variable. For numerical stability the Area was normalised by a factor of 1,000. Inference for a Bayesian model of the form Equation \eqref{eq:5} can be performed via Markov Chain Monte Carlo (MCMC) methods using the function "stan\_mvmer" implemented in the R package ``rstanarm" \citep{Goodrich2020, Muth2018, Gelman2007}. We use default non-informative priors and obtain draws from the posterior distribution of the regression parameters.

\section{Results}\label{sec3}

\subsection{Simulation Study Results}\label{subsec3:1}
\begin{figure}[h]
\centering
\includegraphics[width=0.7\textwidth]{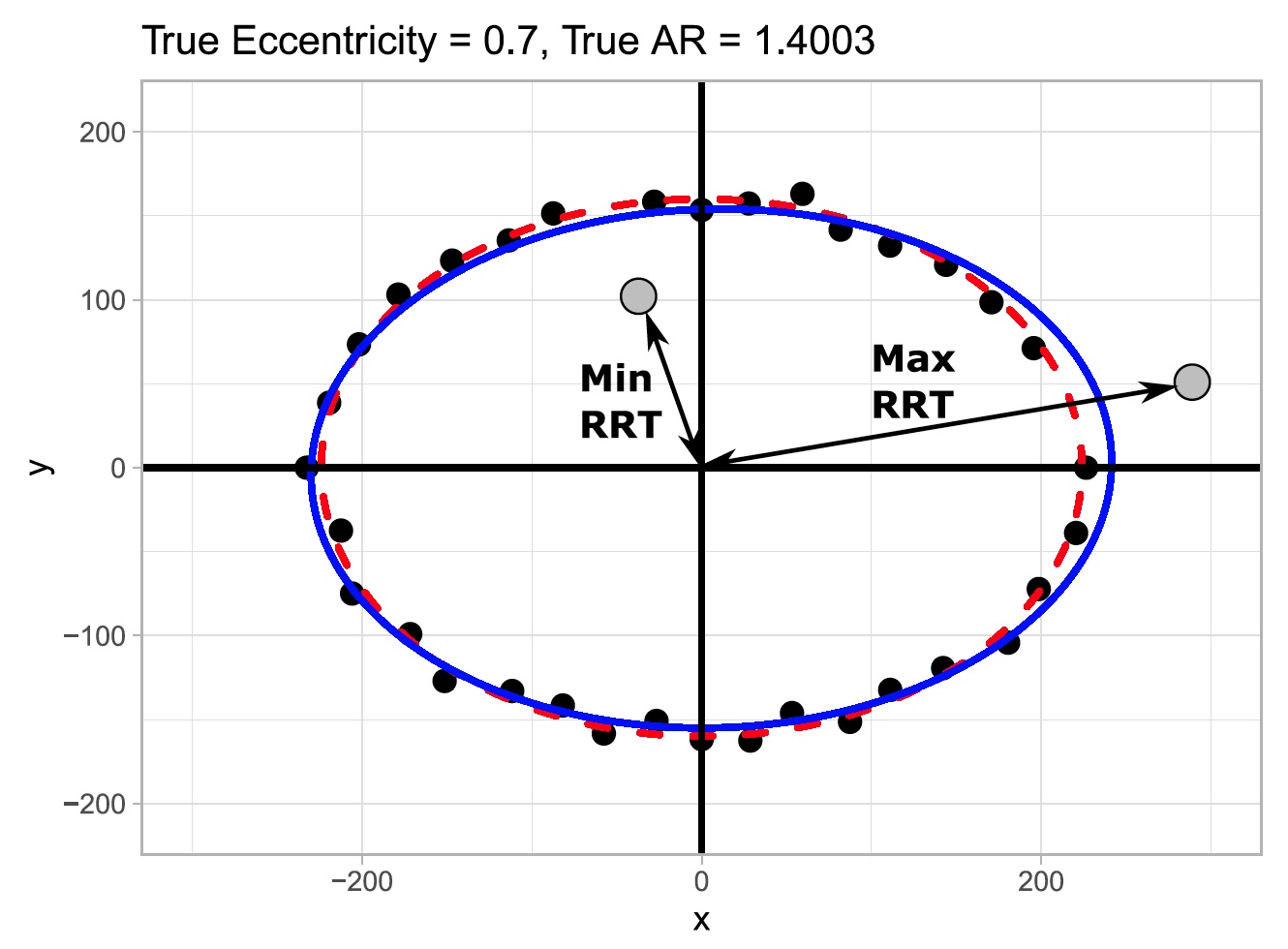} 
\caption{Schematic demonstrating how outliers (in grey) might change the AR dramatically (in this example by increasing the Max RRT and decreasing the Min RRT beyond a range that represents the data) but only slightly elongate the fit ellipse. The red dotted line represents the underlying ellipse with eccentricity 0.7 before noise or outliers were added. The fit ellipse can be seen in blue and has an eccentricity of 0.756. The true AR for the underlying ellipse would be 1.4 however, due to the outliers, the AR in this case would be 2.7.}
\label{fig:4}
\end{figure}

In Section \ref{subsec2:2} we suggested that the eccentricity of a fit ellipse would be a more robust measure of anisotropy than the ratio of the max and min values. An illustrative example in Figure \ref{fig:4} shows how outliers could drastically affect the anisotropy ratio, providing misrepresentative results. However, a more systematic evaluation is required to directly compare the two methods, see Section \ref{subsec2:3}.

\begin{figure}[h]
\centering
\includegraphics[width=0.9\textwidth]{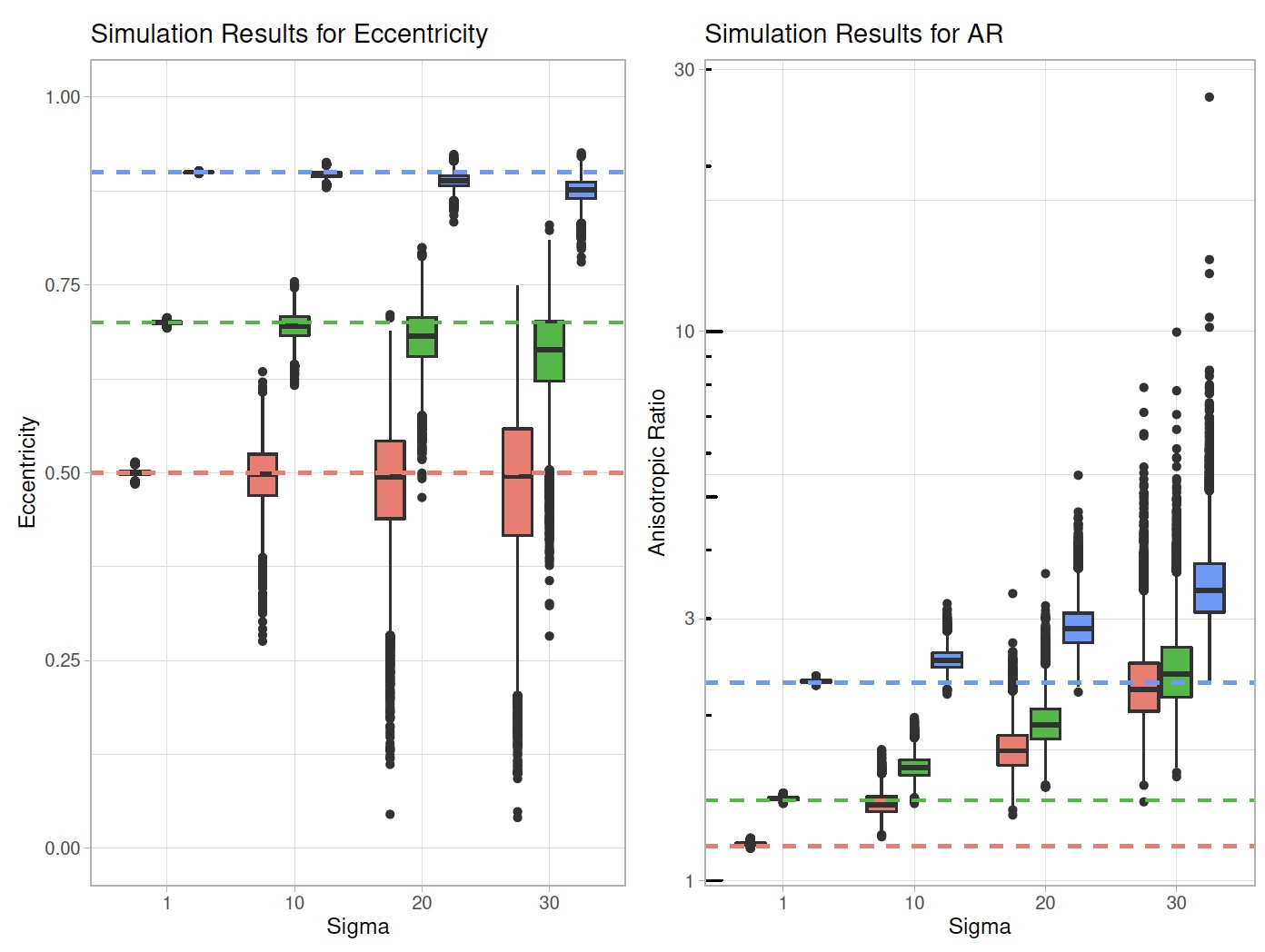} 
\caption{Results of the simulation study for 10,000 observations, $e=[0.5,0.7,0.9]$ and $\sigma=[1,10,20,30]$. The true known value of the measure is represented by the dashed lines. Note that in order to capture the extreme outliers in the AR measure plot (right), the y axis was log transformed. It is clear that for low to high amounts of noise the eccentricity distributions are centered around (or are close to) the true value, whereas the AR distributions stray further and further away from the true value with increasing noise.}
\label{fig:5}
\end{figure}

The results of the simulation study can be seen in Figure \ref{fig:5}. Unsurprisingly, when the noise is very low ($\sigma=1$) both anisotropy measures perform extremely well, they are tightly distributed and centered over the true values. We can see that, as the level of noise increases, the distributions of both measures widen and stray away from the true value. However, we can see that for low to high amounts of noise the eccentricity is far more robust than the AR: the distribution tends to widen but it is still distributed around (or very close to) the true known value. In contrast, as the noise increases, the AR distribution strays further and further away from the true value and the tails of the distribution become more pronounced with extreme outliers appearing. It should also be noted that the AR consistently overestimates the degree of anisotropy, introducing a bias into our measurement, whereas the eccentricity provides a more consistent estimate on average (albeit underestimating the actual eccentricity in some scenarios with high noise). Analysis of particular simulations which resulted in outliers for the eccentricity/AR (where the measure was far from the true value) can be found in Appendix A. From this, and the analysis of the overall distributions, we can conclude that the eccentricity is a more robust and reliable measurement of anisotropy for this type of data.

\subsection{Influence of Age and Gender on skin properties}\label{subsec3:2}
\begin{figure}[h]
\centering
\includegraphics[width=0.9\textwidth]{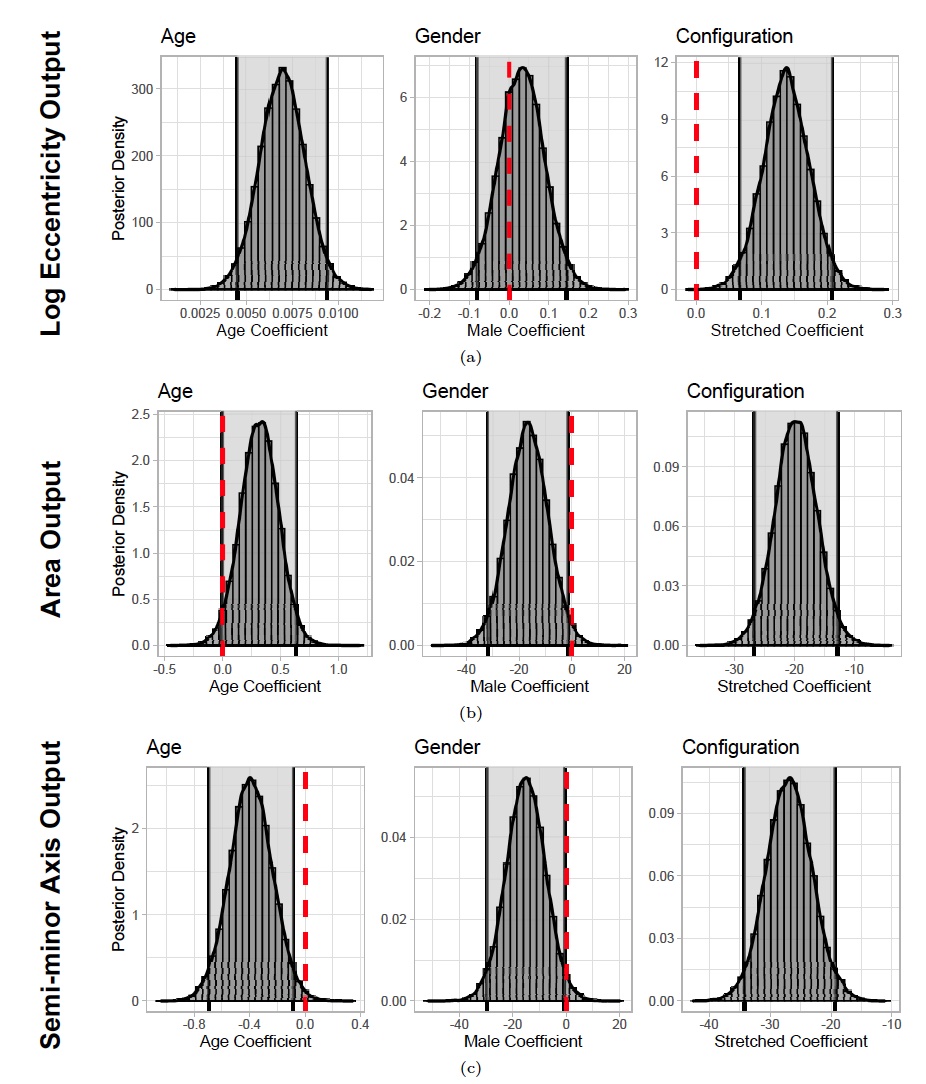} 
\caption{Posterior Distributions for the age, gender and configuration regression coefficients from Equation \eqref{eq:5}. Note that the shaded region between the two vertical black lines represents the 0.95 HPDI and the vertical dashed red line denotes the location of 0. If 0 is within the 0.95 HPDI, there is not enough evidence to say the covariate has a significant effect on the outcome variable.}
\label{fig:6}
\end{figure}

As discussed in Section \ref{subsec2:4}, using a Bayesian approach we can examine the significance of the model covariates by examining the posterior distribution for each parameter. A model of the form Equation \eqref{eq:5} was built and the posterior distributions can be seen in Figure \ref{fig:6}.

\begin{figure}[h]
\centering
\includegraphics[width=0.6\textwidth]{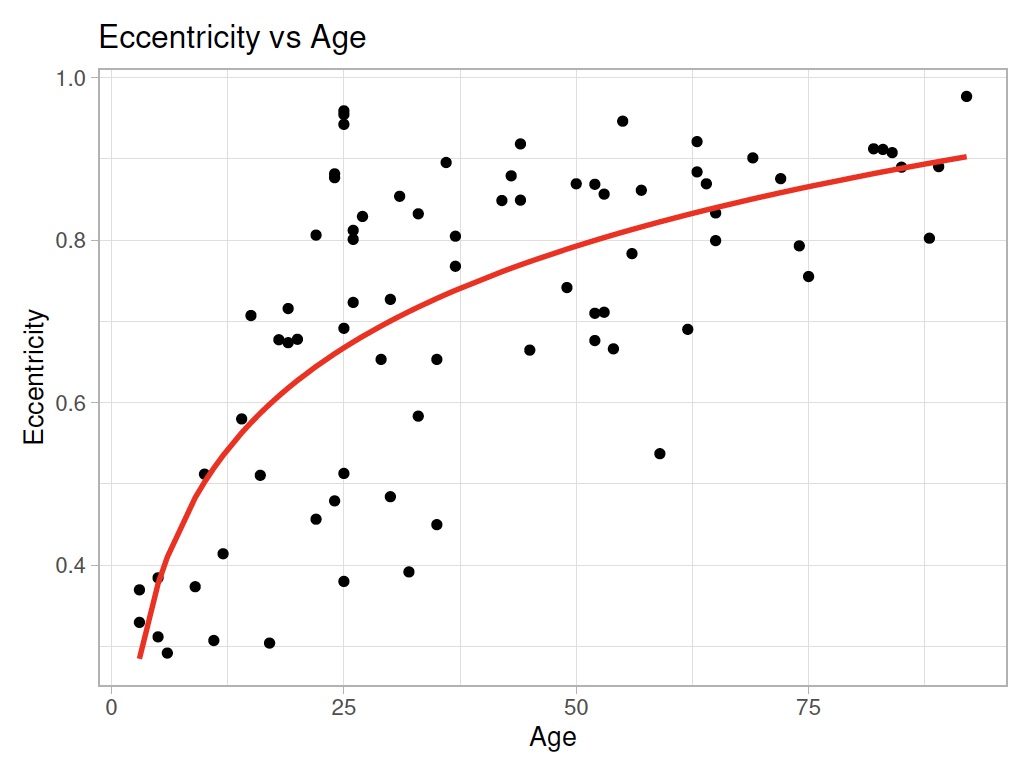} 
\caption{Scatter plot of the eccentricity (i.e. anisotropy) of the fit ellipse vs age of the subject, it is clear that there is a positive relationship, as the age of the subject increases the degree of anisotropy also increases. The red line corresponds to a simple log-linear model fit with $R^2 = 0.5105$.}
\label{fig:7}
\end{figure}

Let's first consider the influence of age. We can see in Figure \ref{fig:6}a that the 0.95 HPDI (grey area) for the age coefficient does not contain 0. Thus, given the data, there is a 95\% probability that age has a significant effect on the degree of anisotropy as measured by the eccentricity. The age coefficient is positive and centred around a posterior mean value of 0.007. Therefore, there is a positive effect of the age on the log of the eccentricity, i.e. the eccentricity increases logarithmically with age. Furthermore, due to the shape of the log function we can conclude that the eccentricity (degree of anisotropy) increases as age increases with a steep increase from childhood into adulthood. This can also be seen by assuming no correlation between the outcome variables and plotting the age of the subject against the eccentricity (see Figure \ref{fig:7}).

In Figure \ref{fig:6}b we can see that the 0.95 HPDI (grey area) barely includes the value zero. Hence, according to the data, age may not have a significant effect on the average stiffness as measured by the ellipse area. However, it is worth noting that a considerable part of the posterior distribution is above the zero threshold, which is in line with the many reports in the literature indicating that skin stiffness increases with age. We must consider that the area here refers to the ``average stiffness" across all directions. In Figure \ref{fig:6}c, we can see the posterior distribution of the effect of age on stiffness in the direction of the Langer line (max stiffness) measured by the length of the semi-minor axis. We can see the age of the subject is significant according to the data. The age coefficient is negative and centered around a posterior mean value of -0.391. Therefore, the data suggests there is a negative effect of age on the length of the semi-minor axis, i.e. as age increases, the length of the semi-minor axis decreases (arrival time of the wave decreases in the direction of the Langer line indicating increased stiffness of the skin in that direction) as expected.

Now, let us consider the influence of gender. We can see in Figure \ref{fig:6}a that the 0.95 HPDI (grey area) for the male coefficient contains 0. Thus, it is unlikely that the gender coefficient has a significant effect on the eccentricity (degree of anisotropy), i.e. according to the data, there is no significant difference between the degree of anisotropy in males versus females. In Figure \ref{fig:6}b we can see that the data shows evidence of a difference between the average stiffness of male subjects versus female subjects. We can also see that, using the category female as the baseline, the coefficient for the shift in area due to the male category is negative and centred around a posterior mean value of -16.725. Therefore, the area of the fit ellipse is on average smaller (average skin stiffness is higher) for males than females. Finally, in Figure \ref{fig:6}c we can see that the data provides evidence for a significant difference between the length of the semi-minor axis (arrival time in the direction of the Langer line) for male and female subjects. The male coefficient is negative and centered around a posterior mean value of -15.174. Therefore, the skin stiffness in the direction of the Langer line is on average higher for males than females.

\begin{table}
\centering
\resizebox{\columnwidth}{!}{%
\begin{tabular}{|l|c|c|c|c|c|c|}
\hline
& \multicolumn{2}{c|}{Age} & \multicolumn{2}{c|}{Gender} & \multicolumn{2}{c|}{Configuration} \\
\cline{2-7}
& Mean & 0.95 HPDI & Mean & 0.95 HPDI & Mean & 0.95 HPDI \\
\hline
log(Ecc.) & 0.007* & [0.005, 0.009] & 0.032 & [-0.081, 0.144] & 0.137* & [0.067, 0.207] \\
Area & 0.316 & [-0.006, 0.636] & -16.725* & [-31.898, -1.546] & -19.832* & [-26.745, -12.864] \\
Semi-maj. & -0.391* & [-0.696, -0.085] & -15.174* & [-29.723, -0.667] & -26.808* & [-34.182, -19.310] \\
\hline
\\
\end{tabular}
}
\caption{Posterior mean values and 0.95 HPDI for the Age Gender and Configuration coefficients where * indicates that the interval does not include the null value 0, suggesting a significant result.}
\label{tab:3}
\end{table}

\subsection{Influence of Skin Tension}\label{subsec3:3}
\begin{figure}[h]
\centering
\includegraphics[width=0.6\textwidth]{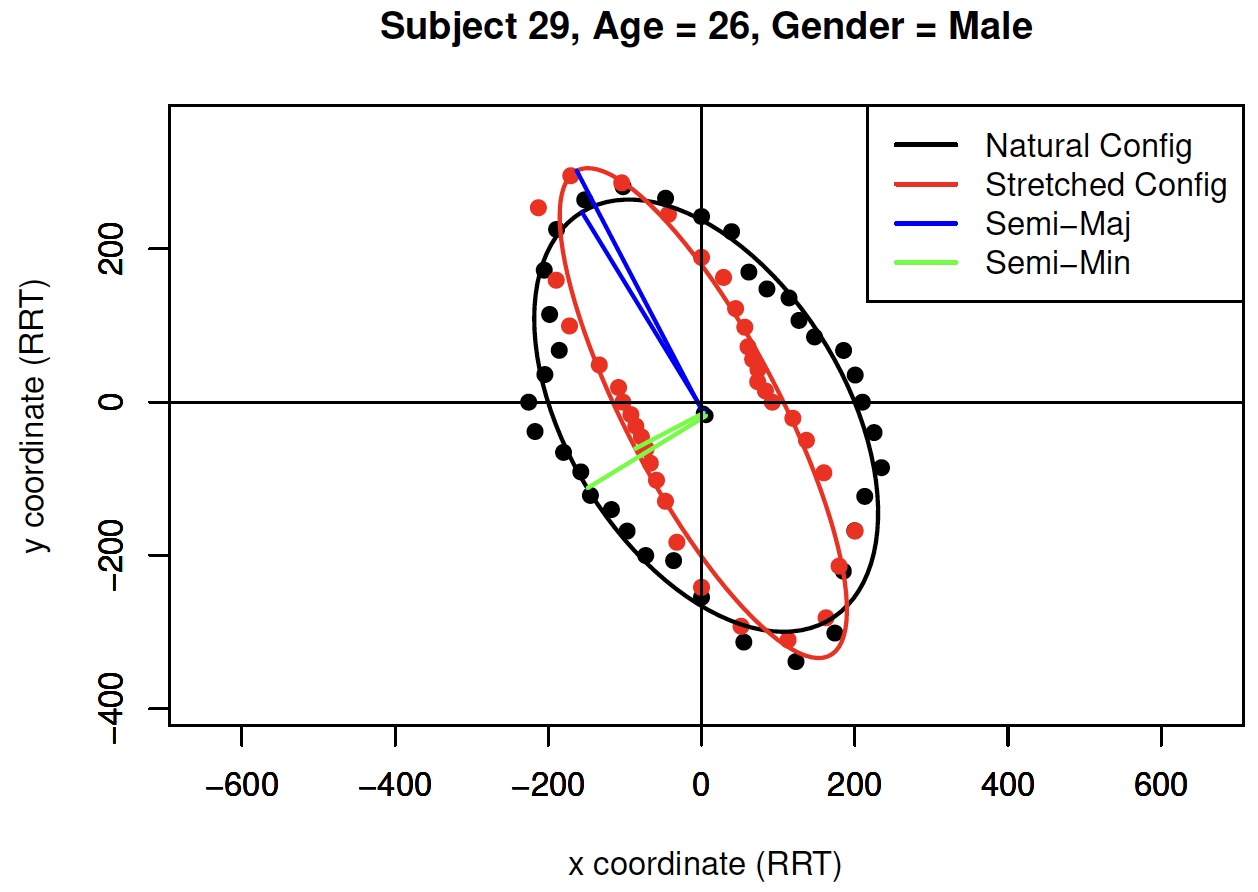} 
\caption{Subject 29, a 26 year old male showing both the natural configuration in black and the stretched configuration in red. Note that the stretched ellipse is narrower and more elongated, but pointing in the same direction, which confirms that the skin tension greatly affects the surface wave speed along the direction of Langer lines.}
\label{fig:8}
\end{figure}

To explore the effect of skin tension on the speed of travelling surface waves and determine the potential efficacy of the Reviscometer in evaluating skin tension, an additional stretch (in the direction of the fastest wave) was manually applied to the skin and the same measurement procedure was carried out. Following this procedure, in theory, we would expect to see:
\begin{enumerate}
    \item The tilt angle should be conserved, provided that the stretch is applied in the direction of the fastest traveling wave (along the Langer line).
    \item The length of the semi-minor axis should decrease due to the increase in wave speed along the direction of the applied stretch.
    \item The eccentricity should increase. We would expect that the stretched data would be more anisotropic due to the additional applied stretch and decrease in semi-minor axis length.
\end{enumerate}
For an example of this behaviour see Figure \ref{fig:8}. Following the same analysis of age and gender in Section \ref{subsec3:2} we can also examine the effect that the additional applied stretch had on the length of the semi-major axis and the eccentricity by examining the posterior distribution of each parameter. Note that as our covariate is a categorical variable with two outcomes (either natural or stretched), the ``stretched coefficient" is the shift in outcome variable from the baseline natural configuration to the stretched configuration. From Figure \ref{fig:6}, we can see that the data provides evidence of a significant difference between the natural and stretched configurations in the log eccentricity and the length of the semi-minor axis. The 0.95 HPDI (grey area) does not contain 0 and the posterior distributions are concentrated far from this value. Furthermore, we can see that the shift for the stretched configuration is positive for the log of the eccentricity outcome (posterior mean of 0.137) and negative for the length of the semi-minor axis (posterior mean of -26.808). This demonstrates that the eccentricity of the fit ellipse increases and the length of the semi-minor axis decreases on average from the natural to the stretched configuration, as expected.

\begin{figure}[h]
\centering
\includegraphics[width=0.6\textwidth]{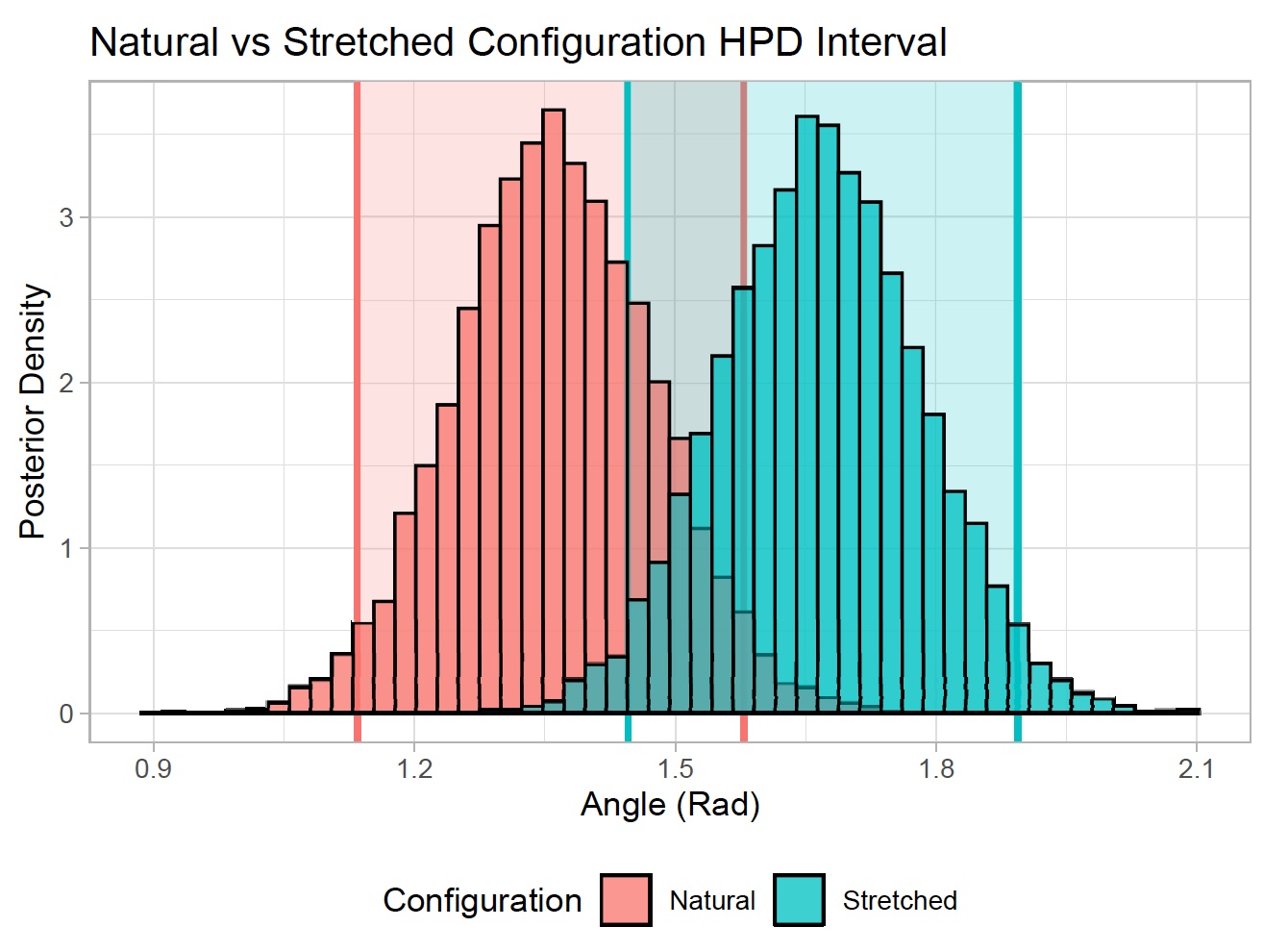} 
\caption{Posterior estimates of the circular means of the angle of the fit ellipse for the natural and stretched configurations. Note that the 0.95 HPDI are represented by the shaded regions which overlap and indicate there is no significant effect of configuration on the angle of the fit ellipse.}
\label{fig:9}
\end{figure}

Finally, to examine the effect that the additional applied stretch had on the angle of the fit ellipse, which indicates the direction of the Langer line, a model was built of the form:
\begin{equation}
\text{Angle of fit ellipse} = \alpha + \beta(\text{Config.}) + \epsilon,
\label{eq:6}
\end{equation}
where $\alpha$ is the intercept, $\beta$ is the coefficient for the configuration and $\epsilon$ is the error. Note that as our covariate is a categorical variable with binary outcomes (natural or stretched), the intercept represents the baseline category (natural) and $\beta$ is the shift in angle for the alternative category (stretched). Note that here we have a circular response variable and cannot build a regular linear model. Following the discussion in \citep{Cremers2018}, we fit a circular regression model using the projected normal distribution within a Bayesian framework, using the MCMC methodology implemented in the function ``bpnr" from the package ``bpnreg" \citep{Cremers2021}. The posterior distributions for the natural and stretched configurations can be seen in Figure \ref{fig:9}. We can see that the 0.95 HPDI overlap for the natural and stretched configurations. Therefore, in the data, there is no indication of a significant difference in the posterior circular means of the two configurations. This indicates that the configuration does not affect the angle of the fit ellipse, hence the direction of the Langer line, as expected. However, there is a degree of uncertainty as the posteriors do not overlap completely so for some subjects there may be variation in angle due to the additional stretch, but the difference in configuration does not appear to be strong in the data.

\section{Discussion}\label{sec4}
As discussed in Section \ref{subsec2:2} we have demonstrated that the eccentricity of a fit ellipse is a better measure of the anisotropy as it is more robust to outliers and noise than the commonly used anisotropic ratio \citep{Ruvolo2007, Boyer2009, Vexler1999}. Robustness to noise and outliers is an important attribute for biological measurements and particularly in vivo skin measurements as we expect experimental error, patient variability and subject movement to impact data collection.

Using our new measure of anisotropy, the relationship between the skin anisotropy and age was explored, see Section \ref{subsec3:2}. It was found that as age increases, the degree of anisotropy also increases, with a steep increase occurring from childhood to adulthood. Many early authors previously reported only a weak dependence of elastic wave speed with age \citep{Dahlgren1984, Vexler1999, HermannsLe2001}, but as noted by \cite{Ruvolo2007}, these studies employed poor resolution angular data (measurements taken every 45 degrees only). More recent studies have noted an increase in anisotropy with age \citep{Thieulin2020, Zahouani2011}, but none of these studies included infants and therefore the steep increase in anisotropy which occurs from childhood into adulthood would not have been captured. \cite{Ruvolo2007} did include infants within their study and report an exponential increase in the anisotropic ratio with age, while in the current study, we have found a logarithmic increase in anisotropy with age. The reason for the different observations may be, in part, due to the use of the anisotropic ratio in \cite{Ruvolo2007} and also due to the fact that participants were divided into 5 age ranges for the purpose of the statistical analysis. In the current study, age was considered a continuous variable and hence provides a richer insight into the true relationship between age and anisotropy. This detailed information can provide evidence as to the validity of applying universal cosmetic surgery practises to very young or elderly patients, where we expect the level of anisotropy to vary significantly.

It has variously been reported in the literature that skin stiffness increases \citep{DulinskaMolak2014, Xin2010}, decreases \citep{HermannsLe2001, Ruvolo2007, Boyer2009, Zahouani2011}, or is not affected at all \citep{Vexler1999} as the age of the subject increases. In Section \ref{subsec3:2}, we concluded that age does not appear to have a significant effect on the area of the fit ellipse (our measure of overall stiffness, independent of direction). However, by virtue of considering the area of the ellipse independent of its shape we are “averaging” over the known fact that the skin of older subjects is more anisotropic. And while the area of the ellipses does not seem to be affected by the age, the length of the semi-minor axis (i.e. max stiffness) does appear to be affected (see Figure \ref{fig:6}c). This indicates that as the age of the subject increases, the stiffness of the skin increases in the direction of the Langer line only. In this context, our results agree with \cite{Thieulin2020} and \cite{Xin2010} who found that there is increased stiffness along the direction of Langer Lines with age. The current results are, however, in direct disagreement with \cite{Ruvolo2007} and \cite{HermannsLe2001} who found that age influenced the maximum RRT (equivalent to the length of the semi major axis in the current study) but not the minimum RRT (equivalent to the length of the semi minor axis in the current study). It is possible that these differences may have arisen due to the method of data analysis e.g. in contrast to the length of the semi-major and semi-minor ellipses fit to the circular data, the use of maximum and minimum RRT values does not account for outliers and does not offer a robust measurement. Contrasting results with other studies may be as a result of reporting “average” stiffness which does not consider the significant anisotropy of skin (in particular elderly skin). We should acknowledge also, a degree of uncertainty with the conclusion that the “average stiffness” (area of ellipse) is not affected by age, as 0 is just inside the 0.95 HPDI and a large portion of the posterior distribution is concentrated above this value (see Figure \ref{fig:6}b). Therefore, perhaps it is unsurprising that there is disagreement in the literature on this point.

Similarly, we must acknowledge a degree of uncertainty with the conclusion that males have stiffer skin than females on average since for both the Area (average stiffness) and the length of the semi-minor axis (arrival time in the direction of the Langer line), 0 is just outside the 0.95 HPDI (See Figure \ref{fig:6}). This result is consistent with \cite{Xin2010} who state that males have stiffer skin (albeit only within certain age ranges, 21-40 years) but contrasts with \cite{Diridollou2000} who state that women have stiffer skin. These contrasting results found in the literature may be as a result of the complex interactions between the influence of age, gender, and body location on the skin stiffness.

On average, the angle of the ellipse is conserved from the natural to the stretched configuration, however, there is quite a wide range of individual behaviours (see Figure \ref{fig:9} and Appendix B). While some extreme behaviours like subjects going from close to $0^{\circ}$ to close to $180^{\circ}$ or vice versa can be explained due to the circular nature of the variable, other variations are likely due to experimental error. As discussed in Section \ref{subsec2}, a fixture was used to take measurements every $10^{\circ}$. Therefore, the identified direction of fastest wave speed (i.e. Langer line) was to the nearest $10^{\circ}$, i.e. an accuracy of $\pm5^{\circ}$. Once the direction of the Langer line was identified, tape was applied manually to stretch the skin in that direction, which also introduces a source of error. This in turn could affect the fundamental mechanics of the probe, introducing further uncertainty, and explaining some of the uncertainty in this result.

During the calibration of the Reviscometer\textsuperscript{\textregistered{}} the material being tested was assumed to be linearly elastic, isotropic and unstressed. While these assumptions are true for the reference materials used (see Table \ref{tab:1}), the application of Equation \eqref{eq:1} to in vivo skin is not strictly valid, because of the pre-tension. Future mathematical relations relating elastic wave speed to stiffness should seek to take this into account. Such a relationship may provide a means to explicitly determine both in vivo skin tension and skin stiffness using elastic waves. A core assumption of this study was that the ellipse is a good fit for our data. We believe this is a reasonable assumption due to the flexibility of an ellipse fit and the data collection procedure which should result in axisymmetric measurements (due to the repeated measurements from $180^{\circ} - 360^{\circ}$) about and perpendicular to the Langer lines. It should be noted however that the raw measurements from some subjects did not exhibit this symmetry and thus the ellipse fit was poor.

In this paper, an in vivo elastic wave technique was employed to investigate the role of age, gender and skin tension on both skin anisotropy and skin stiffness measurements on a sizeable population. By fitting an ellipse to angular data and reporting its eccentricity, we have proposed a more reliable, robust and informative metric of anisotropy than the classic “anisotropic ratio” favoured in the literature. Using a Bayesian approach, we have concluded that skin anisotropy increases logarithmically with age, with a steep increase occurring from childhood into adulthood. Furthermore, the maximum stiffness of skin increases linearly with age, but this increase is only seen along the direction of Langer Lines. We have also concluded that gender does not significantly influence the degree of skin anisotropy, but that both the average skin stiffness, and skin stiffness along the direction of Langer Lines is higher for males than females. Finally, we have also concluded that both the skin anisotropy and skin stiffness measurements are significantly influenced by the level of skin tension present. This suggests that in vivo elastic wave measurements may be a suitable method for inferring in vivo skin tension. To the best of our knowledge, this is the first study which uses a sizeable sample of in vivo subjects and modern Bayesian statistical analysis to evaluate the effect of age and gender on in vivo skin anisotropy. This dataset will provide a useful reference to those wishing to evaluate the effect of subject specific parameters such as age and gender on the anisotropic response of skin.

\section*{Acknowledgments}\label{sec:acks}
This publication has emanated from research supported in part by a Grant from Science Foundation Ireland under Grant number 18/CRT/6049. The opinions, findings and conclusions or recommendations expressed in this material are those of the authors and do not necessarily reflect the views of the Science Foundation Ireland.

\section*{Conflict of Interest}\label{sec:coi}
The authors declare that there is no conflict of interest.

\appendix
\section{Simulation Study - Examination of Poor Performance}\label{sec:appendixA}
\begin{figure}[h]
\centering
\includegraphics[width=0.8\textwidth]{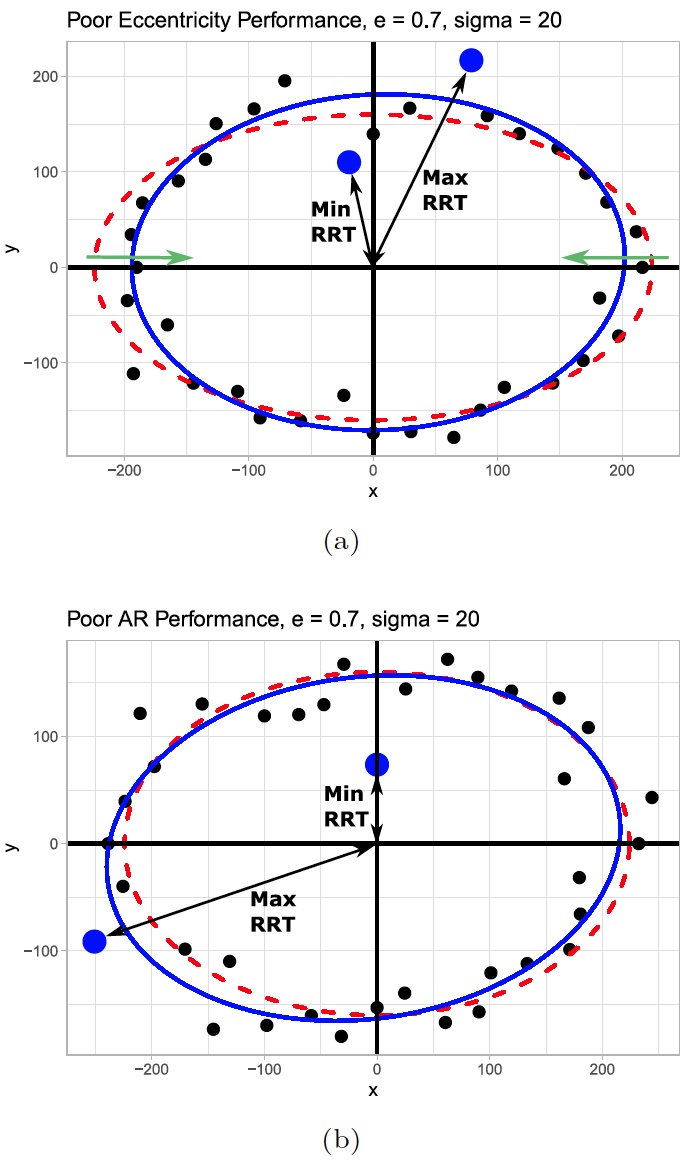} 
\caption{Examination of the conditions under which a measure performs poorly for the scenario $e = 0.7$, $\sigma$ = 20. The red dashed ellipse is the underlying ellipse with true values $e = 0.7$ and $AR = 1.4003$ before noise was added, the blue ellipse is the fit ellipse and the large blue points highlight the Maximum and Minimum RRT values. (a) Case $\hat{{e}}$ = 0.4672, $\widehat{{AR}}$ = 2.0644: The green arrows show the net shift in points by random chance, effectively squashing the ellipse, the blue ellipse appears to be representative of the data despite the low eccentricity. (b) Case $\hat{{e}}$ = 0.7196, $\widehat{{AR}}$ = 3.6236: the ratio considers only the max and min points in blue, disregarding all the other points and performing poorly while the ellipse performs well.}
\label{fig:A1}
\end{figure}

In addition to examining Figure \ref{fig:5}, we can go one step further by analysing examples of poor performance for each measure. For example, let us choose the medium ellipse with medium noise scenario ($e = 0.7$ and $\sigma$ = 20) and examine the performance of both the eccentricity and the AR for two different simulations where:
\begin{enumerate}
    \item Eccentricity is furthest from the true value (i.e. the simulation that produced the largest outlier in the eccentricity distribution).
    \item AR is furthest from the true value (i.e. the simulation that produced the largest outlier in the AR distribution).
\end{enumerate}
This allows us to explore the conditions in which each measure performs poorly and how the other measure performs in those conditions, see Figure \ref{fig:A1}. As we have chosen the middle ellipse, the true eccentricity and AR values are 0.7 and 1.4003 respectively.

In Simulation 668 (Figure \ref{fig:A1}b), the eccentricity of the fit ellipse was 0.7196 and the AR was 3.6236. For this dataset the ellipse performs very well and the poor performance in the AR is driven solely by the two extreme points which are used to calculate the AR but are not representative of the data as a whole. In Simulation 583 (Figure \ref{fig:A1}a), the eccentricity of the fit ellipse was 0.4672 and the AR was 2.0644. For this dataset neither the AR or the eccentricity are close to their true values, the AR significantly overestimates the degree of anisotropy using two extreme values that are only 30 degrees apart (rather than 90 degrees which would make physically intuitive sense). The poor performance in the eccentricity is because, by chance, the random noise pulled the points along the semi-major axis closer to the origin and pushed the points along the semi-minor axis away from the origin. Thus, even though we know the “true” eccentricity before noise to be 0.7 it could be argued that the calculated eccentricity of 0.4672 is a better measure of the anisotropy as it is representative of the data.

This, along with the analysis in Section \ref{subsec3:1} allows us to conclude that eccentricity is a more robust measurement of anisotropy for this type of data.

\section{Individual Subject Behaviours}\label{sec:appendixB}
\begin{figure}[h]
\centering
\includegraphics[width=0.5\textwidth]{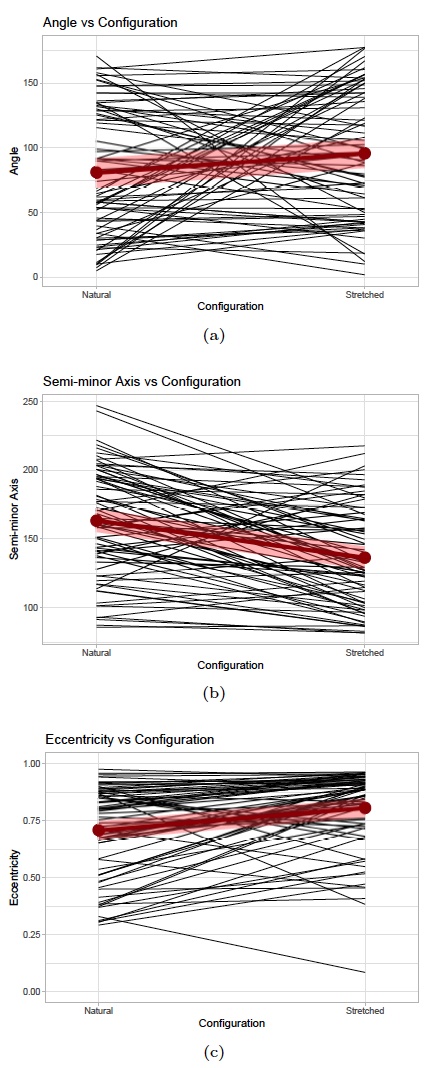} 
\caption{Spaghetti plots where the red circles denote the mean and the light red region is the 95\% confidence interval of the mean. (a) There is a wide variety of behaviours for the angle but on average the angle remains constant. Note that some subjects go from close to 0 to close to 180 or vice versa, this behaviour is equivalent to a small increase/decrease in angle. (b) There is some variance in behaviour but the semi-minor axis decreases for the majority of subjects. (c) There is some variance in behaviour but the eccentricity increases for the majority of subjects from the natural to the stretched configuration.}
\label{fig:B2}
\end{figure}

As well as looking at the average behaviours in Figure \ref{fig:6} and Figure \ref{fig:9} we can examine the specific behaviour of individual subjects using spaghetti plots (see Figure \ref{fig:B2}). Most subjects exhibit the expected behaviour for the length of the semi-major axis and the eccentricity with only a small minority of subjects that have alternative behaviours (see Figures \ref{fig:6}a and \ref{fig:6}c). For the angle, we can see quite a range of different behaviours (see Figure \ref{fig:B2}a) but on average, most subject angles remain more or less the same. Note that some subjects go from close to $0^{\circ}$ to close to $180^{\circ}$ or vice versa, this behaviour is equivalent to a small increase/decrease in angle due to the circular nature of the measurement.

\bibliography{references}%
\end{document}